# Investigating a Conceptual Construct for Software Context

Diana Kirk
School of Computer & Mathematical
Sciences Auckland University of Technology
Private Bag 92006, Auckland 1142, New
Zealand diana.kirk@aut.ac.nz

Stephen G. MacDonell
Department of Information Science
University of Otago
PO Box 56, Dunedin 9054, New Zealand
stephen.macdonell@otago.ac.nz

**Abstract**

*A growing number of empirical software engineering researchers suggest that a complementary focus on theory is required if the discipline is to mature. A first step in theory-building involves the establishment of suitable theoretical constructs. For researchers studying software projects, the lack of a theoretical construct for context is problematic for both experimentation and effort estimation. For experiments, insufficiently understood contextual factors confound results, and for estimation, unstated contextual factors affect estimation reliability. We have earlier proposed a framework that we suggest may be suitable as a construct for context i.e. rep- resents a minimal, spanning set for the space of software contexts. The framework has six dimensions, described as Who, Where, What, When, How and Why. In this paper, we report the outcomes of a pilot study to test its suitability by categorising contextual factors from the software engineering literature into the framework. We found that one of the dimensions, Why, does not represent context, but rather is associated with objectives. We also identified some factors that do not clearly fit into the framework and require further investigation. Our contributions are the pursuing of a theoretical approach to understanding software context, the initial establishment and evaluation of a construct for context and the exposure of a lack of clarity of meaning in many 'contexts' currently applied as factors for estimating project outcomes.*

## 1. INTRODUCTION

In the domain of software engineering, evidence suggests that practitioners adapt development methodologies to suit specific project contexts [2, 3, 17, 22, 32, 33, 36, 15, 42]. Moreover, research indicates that most organisations adapt practices from several approaches, often at the level of the individual project. As an example, as agile approaches have become more established, deficiencies have been exposed, leading to either contextualisation [23] or to amalgamation with other paradigms, for example, the 'lean' paradigm [45]. MacCormack et al. suggest that firms must deploy different business processes according to business context and that applying a uniform 'best practice' approach results in missed opportunities [32]. In addition to the issue of tailoring, the emergence of new paradigms has highlighted limitations in current approaches. Those researching software projects in a global environment suggest current practices must be ex- tended or modified to suit this specific environment [37, 43]. The software-as-a-service paradigm has raised new possibilities for evaluation of user behaviour and new kinds of approach are required as current methodologies do not satisfy the needs of these new opportunities [41]. The traditional viewpoint — that methodologies and practices should be adopted and used as prescribed [14] — has thus been superseded by one of acceptance that tailoring according to project-specific contexts is both necessary and unavoidable.

This state of affairs raises questions about software context. If context is of importance for software practice selection, we clearly must strive to fully understand the nature of the relationships between practice and context. Only then will we be in a position to advise industry on which practices might be most suitable or to predict project out- comes based on context and practices implemented. The sole route to this kind of general understanding is through theory-building [19]. Without an understanding of the relationships between context, practice and outcomes, we can at best expose patterns based on data or observations. Such patterns represent correlations and *correlations* can not be used to predict in a general way. Basili et al. remind us that, when carrying out controlled experiments, "... it is hard to know how to abstract important knowledge with- out a framework for relating the studies" [5]. A theory is a model or framework for understanding.

The role of theory in software engineering (SE) has been investigated from a number of perspectives. Sjøberg et al. observed that there is very little focus on theories in software engineering and remind us of the key role played by theory- building if we wish to accumulate knowledge that may be used in a wide range of settings [38]. Hannay et al. con- ducted a review of the literature on experiments in SE and found that fewer than a third of studies applied theory to explain the cause-and-effect relationship(s) under investigation, and that a third of the theories applied were proposed by the article authors [21]. Gregor examined the nature of theory in Information Systems and found multiple views on what constitutes a theory [20]. Her investigations resulted in a taxonomy of theory types along with a proposed list of structural components of a theory. Stol and Fitzgerald argue that SE research does, in fact, exhibit traces of theory and suggest that the current focus on evidence based soft- ware engineering (EBSE) must be combined with a theory-focussed research approach to support explanation and prediction [40].

While there is general agreement within the research community that an increased focus on theory building would produce benefits, there remains uncertainty about how to proceed. One of the structural elements of a theory as pro- posed by Gregor is the construct and she states that "All of the primary constructs in the theory should be well-defined" [20]. We suggest that the area of software context represents one example of how lack of a defined construct creates problems for researchers and practitioners. In the first instance, researchers who carry out formal experiments are unable to confidently interpret the scope of applicability of their results because the role of contextual factors is insufficiently understood [5, 7, 30, 39]. Second, there is inherent uncertainty in the use of available data repositories for investigation and estimation (for example, estimation of project effort) because the environment associated with the data is at best only partially stated [6].

Our overall goal is to support practitioners in the selection of practices appropriate for specific contexts. We have earlier suggested that, for software, achievement of this goal requires a deeper understanding of the relationships *between business objective, process* and *context* [29, 28]. It is our viewpoint that these must be defined and operationalised before meaningful progress can be made. We have proposed a six-dimensional framework for context, with dimensions described as *Who, Where, What, When, How* and *Why* [27]. Our strategy for testing this framework for suitability as a theoretical construct for context is to categorise contextual factors from the software engineering literature into the framework. Success in this would provide us with some confidence that the framework provides a structure that will al- low us to view the space of all contexts in a simpler and thus more manageable way. Our vision is that, as researchers understand the kinds of information that need to be captured as 'context', growing evidence repositories will yield 'practice families' i.e. sets of similar practices that are indicated (and contra-indicated) for a specific value along one dimension. 'Best practice' for an initiative will then involve choosing from practices that are consistent with values along all dimensions i.e. either indicated or not-contra-indicated.

Our research objective for this paper is to present the results of a pilot study in which a small number of sources were used as proof-of-concept. Our contributions are the pursuing of a theoretical approach to understanding software context and the introduction of a construct for context. In section 2, we consider existing approaches to theory-building related to software and overview studies aimed at categorising contextual factors. In section 3, we present our framework. In section 4 we present our research approach and in section 5 we discuss findings. In section 6, we summarise the paper and discuss limitations and future work.

## 2. RELATED WORK

There are two areas of related work for this paper. The first includes attempts to provide suitable theoretical constructs for the problem space of software development. The second includes research aimed at more informal efforts to categorise context along various dimensions.

**2.1 SE Theory Building**
As introduced briefly above, Sjøberg et al. remind us that "building theories is the principal method of acquiring and accumulating knowledge that may be used in a wide range of settings" [38]. The authors describe the building blocks of a theory as including constructs (the entities in the domain), *propositions* (how the constructs are related), *explanations* as to why these relationships hold and the scope in which the theory applies. Of the initiatives aimed at theory-building in software engineering [38, 48], the issue of context has been treated as secondary i.e. none (to our knowledge) has ad- dressed the need for a deeper understanding of context. For example, Sjøberg et al. propose a framework that includes *Software system*, which "may be classified along many dimensions, such as size, complexity, application domain, ...". The form of possible classifications for context is not discussed.

Taking a different approach, Dybå et al. consider the distinction between omnibus and discrete context, where the former "refers to a broad perspective" and the latter to "specific contextual variables" [16]. The authors observe that most empirical SE research to date has adopted the discrete perspective and state the need for the broader perspective. As in our approach, such a perspective involves a consideration of *Who, Where, What, When, How* and *Why*. However, the dimensions and their meanings (adapted from organisational science) are presented as a given, and the authors do not pursue the link between the ideas of omnibus context and a theoretical construct for software context.

An important current endeavour that aims to develop a general theory for software development is the SEMAT initiative, launched in 2009 by Ivar Jacobson, Bertrand Meyer and Richard Soley [24]. The initiative is backed by many of high rank within the software discipline. The approach proposes a SEMAT kernel, a set of elements that name the basic concepts. The kernel comprises seven top-level ideas - *Requirements, Software System*, Work, Team, Way of Working, *Opportunity* and *Stakeholders*. These concepts support determination of a project's health and facilitate selection of a suitable set of practices.

Although the authors of the SEMAT approach state that the initiative promotes a "non-prescriptive, value-based" philosophy that encourages selection of practices according to context [25], we suggest that the approach is, in fact, prescriptive in intent. Each of the SEMAT elements has a number of states "which may be used to measure progress and health" [25], and this implies that the health of every software project can be measured in a common way i.e. the measurement of health is prescribed. We believe that the variation in software project objectives along with the people-centric nature of software development is not consistent with the idea of a pre-defined notion of 'health'. One last issue with the SEMAT initiative relates to the claim that a general theory is being developed. Weiringa, discussing the field of requirements engineering, reminds us of the dangers of confusing solution design with research [46]. We suggest that SEMAT, at this point, represents a design initiative rather than a theory building exercise.

### 2.2 Factors Based Approaches

There have been many efforts to relate project outcomes to specific key project factors. We overview a selection here. Avison and Pries-Heje aimed to support selection of a suitable methodology that is project-specific [2]. For a given project, the authors plotted position along each of eight dimensions on a radar graph and inferred an appropriate methodology from the shape of the plotted graph. While we support the intent to understand context in this way, we see two limitations in the work. First, the abstracted categories are based on inputs from a single organisation and so are inevitably scoped to the operating space for the organisation. This means that, although key ideas such as quality and culture are included, some important contexts are missing, for example, temporal distance. A second limitation is that the abstraction is based at the level of the project. In the realm of software, this has im- plications of spanning requirements determination through to product delivery and we suggest that this scope is not sufficiently broad. For example, the most important practices recommended for new product development are at 'the leading edge' i.e. involve issues of strategy and product de- termination, and these occur before a 'project' to develop the new product is commenced. Another example involves the 'software-as-a-service (SaaS)' delivery paradigm. Here we note that the emphasis changes from a 'developer driven' to a 'customer-driven' environment, where the on-going relationship between development group and customer becomes key [41]. Again, this is not part of a 'project' as it is generally understood. Clarke and O'Connor propose a reference framework for situational factors affecting software development [9]. One aim is to "develop a profile of the situational characteristics of a software development setting" and to use this to support process definition and optimisation. The framework includes eight classifications: Personnel, Requirements, Application, Technology, Organisation, Operation, Management and Business, further divided into 44 factors. Our critique of this approach is that it remains 'factors-based'. Although factors are grouped into classifications, there is no meaning that helps us understand relevance. For example, the factor 'Cohesion' represents a number of different kinds of 'meaning', referring to "team members who have not worked for you", "ability to work with uncertain objectives" and "team geographically distant". These three meanings can be viewed as quite different.

Kruchten presents a contextual model based on experience for situating agile practices. The aim of the model is to "guide the adoption and adaptation of agile development practices", particulary in contexts that are "outside of the agile sweet spot" [31]. The model is interesting to those committed to an assumption of 'agile practices as basis but may be adapted'. However, it is situated in the solution space of agile projects. Wang notes that the need to align practices with business goals is a limitation of agile practices which "do not generally concern themselves with the surrounding business context in which the software development is taking place" [45]. As for the Avison and Pries-Heje study above, this model excludes some important contexts and has narrower scope than we require.

Orlikowski carried out an exploration of a globally-dispersed, multinational product development organisation, and observed a number of boundaries that served to shape and challenge the distributed product development [34]. These boundaries are *temporal* (multiple time zones), *geographical* (multiple global locations), *social* (many participants engaged in joint development work), *cultural* (multiple nationalities), *historical* (different product versions), *technical* (complex system, various infrastructures, variety of standards) and *political* (different interests, local versus global priorities). In Orlikowski's research, the problem space is described. However, the work does not relate specifically to software organisations and the criticism of 'single organisation' remains. Aspects such as business goals and the ongoing relationship between customer(s) and development group are not included. This work, however, informed the initial basis of our context framework.

Zachman created a 2-dimensional framework for describing an enterprise architecture [47]. The first framework dimension comprises the columns *Who, Where, What, When, How and Why*, and the second dimension contains a number of perspectives on the organisation. Although the meanings of the first dimension columns are not useful to our research (for example, the meaning of What as a list or model of artifacts does not support practice categorisation), we recognise that the column categories map to the dimensions of our context framework [27].

### 3. FRAMEWORK FOR CONTEXT

Our viewpoint is that we must provide a decision support mechanism based on our understanding of objectives and situated process that supports the practitioner in making decisions regarding practice selection. We begin from the idea that our task is to understand the problem space, including context, in a way that supports such decision-making. As part of our journey towards better understanding software context, we have created a framework with semantics based principally on the works of Orlikowski [34], and with syntact mapping to the models of Zachman [47] and Dybå et al. [16].

We visualise that our framework will be of use to both researchers and practitioners. At present, researchers studying the effects of specific practices on project outcomes tend to capture context in an ad-hoc way. For example, the experience level of participants might be noted as believed to be of relevance. However, there may be other aspects of context that also affect practice efficacy. For example, it may be that formal inspections are inappropriate under conditions of new product development, regardless of practitioner experience. Until we understand all relevant aspects, we will fail to understand the operating limits for a practice and will be unable to advise industry. On the other hand, once we have understood operating limits, we may apply the framework to produce a 'blueprint' for an organisation or project, and may then advise on practice suitability, based on available evidence. Of course, there are many possible contexts and we suggest that one consequence of this is that it will not be possible for the research community to produce a prescriptive 'solution' that covers all possibilities. Rather the aim is to provide decision support, where a list of suitable practices is presented to the practitioner based on some key aspects of context, allowing the practitioner to make the final selection based on local knowledge.

| Dimension | Examples |
|---|---|
| Who | Consistency in world views: affected by nationalities, culture, team structure, power structures, etc. |
| Where | Physical distance; temporal, locational. |
| What | Product-related constraints: affected by standards, external product interfaces, required quality, etc. |
| When | Life-cycle stage of the situated product. |
| How | Engagement constraints: affected by client delivery expectations, expected involvement, etc. |
| Why | Organisational drivers: result in strategies that cause constraints in the other 5 dimensions. |

Figure 1: Model dimensions

We believe that this approach might also help expose potential issues in advance. For example, a recent study of situated agile practices uncovered an issue concerning the need for detailed documentation to comply with certification requirements. Some developers perceived documentation as being anti-agile, and this caused problems of internal team clashes [23]. We suggest that, had a suitable frame- work for evaluating context been available, one that took into account the need for standards, the evangelism of some team members may have been exposed and dealt with in advance.

The proposed framework has six dimensions *Who, Where, What, When, How* and *Why*, each dimension representing a key aspect of software context [27]. In Figure 1, we overview the meanings of these aspects along with some examples and initial thoughts on dimension structure.

**Culture (Who)** The world-views of stakeholders will influence which practices are likely to be successful. Issues of power, language and expectations about 'how things work' will be relevant. This dimension is manifested in how people are structured to carry out work. A group within a larger organisation may be constrained to adopt the processes of the larger group. A management team and development team will probably have different world-views. A group that likes and expects change might want to experiment with new technologies. A team of developers that works on several projects at the same time differs culturally from a team that is dedicated to a single project. As there are many aspects that affect an individual's view of the world, we expect this dimension to be extremely 'rich'. It is not at this point clear what an appropriate structure might look like. However, we believe a suitable starting point is the layered behavioural model provided by Curtis et al. in 1988 [13, 12]. The five layers provided by this model include, for example, 'Individual', 'Team' and 'Company', the suggestion being that different kinds of behavioural analysis are appropriate for each layer. This may provide a basis for a structuring for this dimension.

**Space and time (Where)** How stakeholders are separated in space and time will place constraints upon practices relating to communication and co-ordination. The separation might affect any team interface and so this dimension has many possible scenarios. Some examples are remote clients, outsourcing testing and non-colocated teams. A naive first-level structure for this dimension includes 'separation in space' and 'separation in time'. For the former, the notion of interfaces between groups may be relevant i.e. which groups are separated? For both, the degree of separation seems relevant i.e. 'how far apart in space or time'. However, Clear et al. suggest that notions of time [11] and space are complex and propose a model for spacial distance that takes into account the mobility of team members [10]. A structure for this dimension may include a first-level separation into 'space' and 'time' with a second-level structure that takes mobility and other notions into account.

**Product constraints (What)** The nature of the software product may require consideration of standards, product external interfaces and quality expectations. Software for the chemical industry may require adherence to specific process and/or product standards. Embedded software, middleware and stand-alone applications involve different kinds of product interface and possibly different practices relating to product integration. Software for medical equipment may require formal quality practices to be in place. From the examples identified, we surmise that structure might include 'ap- plication subject area' (e.g. chemical or medical equipment) and 'product relationship' (e.g. stand-alone, embedded or middleware).

**Product life-cycle stage (When)** A situated product moves through a tailored life-cycle that includes, for example, initial creation, adolescence, maturity and retirement. Notions of 'health' and the practices that affect outcomes will be different at various points in the cycle [26]. For example, much of the new product development (NPD) literature indicates that practices such as clarification of NPD goals, identification of value proposition, cross-functional teams and go/no-go gates are more important than 'in-development' practices for successful outcomes. Later in the cycle, when the product is in use and there are many requests for new functionality from different sources, change management practices become key. Still later, when the product is nearing retirement, it is likely that practices to keep 'important' customers happy may be most important, during a transition to a newer, more relevant product. MacCormack et al. discuss three life-stage related contexts: new product start-up, best approached by development practices that support emergence; product growth, which requires an agile approach for managing rapid product evolution; and product maturity, which requires efficient processes that reduce costs [32]. Furneaux and Wade discuss product discontinuance and provide a model to support the formation of clear discontinuance intentions as a prelude to decision-making [18]. At this stage, we envision this dimension as having a linear structure that represents the life stage of the situated product.

**Engagement constraints (How)** The demographic of the receivers of the software product or service may influence product specification and delivery mechanisms. Delivering custom software to a single customer prob- ably indicates practices such as evolutionary specification, many deliveries, prototyping and customer involvement. Deploying a telephony middleware product to a vertical market with different feature expectations may require a well-defined roadmap, product-line approach, strong customer relationships, effective pre-delivery testing and a defined and infrequent delivery cycle (as the receiving organisations may need to ratify middleware within their own systems). Deploying to many probably requires greater focus on requirements and change management. An early adopter market may expect fast delivery with frequent updates. A structure for this dimension might include 'specification constraints' and 'delivery constraints', with both sets of constraints resulting from client-internal processes. Note that these constraints are different from those imposed by culture (who) or distance (where).

**Organisational drivers (Why)** The key strategic goals of the producer organisation will affect practice suitability. A goal of 'reduce time-to-market' suggests practices that support speed rather than quality. A goal of 'go global' may exclude practices that require heavy involvement of end-users.

The thesis is that what constitutes 'best practice' for a project will depend on the project's profile with respect to the above dimensions. Indicated practices for a single dimension are those which have been found to be effective for the profile value in this dimension. Indicated practices overall are those that exist at the intersection of all dimensions. For example, 'pair programming' may be indicated from cultural and locational perspectives, but may be contra-indicated for an 'early-adopter' demographic, where fast delivery is expected over high quality.

Clearly, there are many possibilities for assigning meaning to the dimensions 'Who', 'Where', etc. Our rationale for assigning meaning as above is based on the idea that different dimensions relate to different kinds of activity. In particular:

**Who** associates with *understanding how things work*.

**Where** associates with *communicating and co-ordinating*.

**What** associates with *defining the product*.

**When** associates with *understanding the situated product*.

**How** associates with *establishing engagement rules*.

**Why** associates with *establishing objectives*.

## 4. RESEARCH APPROACH

The research question we address in our current research is Does the proposed dimensional framework represent a suitable construct for the space of software contexts? Specific sub-questions are:

**RQ1** Can all contexts in software projects be categorised as belonging to a framework dimension?

**RQ2** Can each context be categorised as belonging to only one framework dimension?

**RQ3** What is the internal structure of each of the framework dimensions?

RQ1 and RQ2 address the questions of completeness and orthogonality. RQ3 addresses construct complexity — if overly complex, the framework will be impractical as a research construct and so probably not useful in practice.

Our approach was to identify contextual factors from the literature and then attempt to categorise each found factor as one or more framework dimension. Before proceeding with a full search, we carried out a pilot study, in which we selected and analysed three documents, obtained by applying the search strings to one of the search sources (see below). (While three documents is clearly a very small initial sample, the richness of information drawn from them with respect to context, as discussed in the next section, suggests that this number was suitable in supporting this step in our research.)

Before proceeding with the search, we clarified what we meant by 'context'. A suitable accepted definition of the term is "The circumstances that form a setting for an event, statement, or idea, and in terms of which it can be fully understood" [35]. We adapt this for software projects as "The circumstances that form a setting for an organisational initiative, and in terms of which outcomes can be fully under- stood" i.e. for any single initiative, such as a 'project', we must identify the set of circumstances that is relevant for understanding the outcomes for that initiative.

Based on this definition, we define context as having two aspects:

- Any factor that affects how well a practice meets objectives.
- The factor cannot be changed i.e. represents a hard constraint. For example, specific individuals may have been allocated to a project and the project manager can not alter the allocation, or an organisation has a policy that all acceptance testing will be carried out by a remote, dedicated test team.

Our definition results in the notion that context for one initiative or project may not be context for another. For example, physical location of team members may be fixed in one instance (hard constraint) but may be changeable in another (soft constraint, and so not part of context). Our definition also implies that business objectives and implemented techniques and tools are not part of context. The former, although highly important for practice selection, represents a measure of success for a practice (how well did the practice meet the objective) rather than a context that affects how likely it is that the practice will be effective. Techniques and tools are inherently part of implemented practices and so again do not represent context.

**4.1 Search Strategy**

For the full study, we plan to source titles and abstracts from each of:

- Elsevier's Scopus for IS technical and social sciences literature
- Academic Search Premier (EBSCO) for business focus
- Google Scholar to explore wide range of sources

The pilot that is the subject of this paper involved searching the first source only (Scopus). Our search string was ("software development" OR "software engineering" OR "soft- ware process" OR "software project" OR "software study" OR "software management") AND (context OR factor) AND (outcome OR success OR failure)). This resulted in 2011 documents.

**4.2 Selection Strategy**

As this study involves testing a general framework, we must be as comprehensive as possible in our identification of contextual factors. This means that we want to expose factors that may not be considered as such. For example, in Section 1, we noted that the software-as-a-service paradigm has exposed a need for different kinds of practice, but this is not normally viewed as a contextual factor. We thus would like to include studies that contain any thoughts or description about what might affect practice efficacy.

We accepted documents that described studies relating to situated process or practice. We rejected studies that:

- did not relate to software organisations or software projects
- focussed on process efficacy, IT solutions adoption, products, techniques, metrics or tools
- related to factors for software process improvement
- were not situated in industry

From the accepted documents, we extracted words or terms that could be viewed as stating or describing a contextual factor. We refer to these words or phrases as elements. As our aim was to ensure broad coverage, we did not evaluate the studies in which the elements were mentioned for quality i.e. we accepted a source if it mentioned an element that was perceived or reported by the author(s) as being associated with project outcomes. We did not consider the nature of the constraint. We did not 'tidy up' the found elements by making value judgements about whether two elements had the same meaning. We felt that such evaluations would effectively remove some of the nuances of identification and would thus weaken the evaluation of the framework. The underlying issue here is one of a lack of common, agreed vocabulary for software projects.

A first pass selection of articles from Scopus was carried out by the first author based on title and abstract to create a list of candidate primary studies. These were sorted according to publication date (newest at the top) and then first author surname. For this pilot, a second pass was then carried out by the first author on the full text of the first four papers in the resulting candidate list. One was rejected as the full text was not easily accessible, leaving three suitable for inclusion [1, 4, 8].

# 5. FINDINGS

In Section 4, above, we noted that our definition for Con- text implies that business objectives are not part of con- text. This means that the dimension Why, as defined for our framework, can not be included in a construct for Con- text. On further thought, this makes sense. For example, a goal of 'corner the market' may be addressed with a strategy of new product innovation (when), fast delivery (how) and involving customers in development (who). The goal affects strategic decisions which, in turn, affect practice suitability i.e. the goal is not directly related to practices. Our proposed framework is reduced to five dimensions.

In Figure 2, we show some of the extracted context elements for the dimensions Who, Where and What (for the three papers studied, these three dimensions were the 'richest' i.e. contained the largest numbers of terms). We also show elements that did not obviously map to any of the proposed dimensions.

The first observation is that there are many elements for Who. It is not clear whether this is the result of the small sample containing a study relating to perception gaps (essentially cultural in nature) or whether this dimension will become too complex to be useful. A more exhaustive study will clarify this. We also comment that, for such a small number of sources, a large set of possible contextual elements has been identified. This in itself is an interesting discovery as we might expect the context space we will un- cover in the study proper will be extremely rich, and provide data for analysis of the structures of the individual dimensions.

However, we note that, although most found elements were easily categorised into a single proposed dimension, there were some that we could not obviously categorise. Some of these we suggest can be removed. For example, 'Strategy' can probably be removed, as we have noted above that this will determine decisions affecting dimensions i.e. it is not a contextual element in its own right. 'Company size' may also be a candidate for removal, if we hypothesise that size in itself will not affect outcomes, but rather what this means in terms of culture and physical and temporal separation. Such a hypothesis would imply that past studies in estimation model-building may have been focussing on factors that are 'secondary' i.e. comprise a number of more basic elements. This in turn implies the possibility of a more refined approach to estimation in the future.

We next consider more deeply each of the unmapped elements with a view to understanding for each whether it requires an extension to the framework at top level i.e. an additional dimension, whether it serves to improve our understanding of possible structure for a dimension, or whether we can make a case for treating it as 'secondary'.

| Who | Where |
|---|---|
| *Developer* | Company in one country; development in two other countries. |
| Intention to adhere to specific practices. | Cross-functional teams. |
| Understanding of, and expertise in, problem domain. | Distributed dynamic teams. |
| Professional competence. | Insourcing/outsourcing (work done by employees) |
| Technical competence. | On-shore/off-shore(work done in home country of the organisation) |
| Attitude caused by perceptions of age and gender . | Subject matter experts, analysts and project managers at each site. |
| Specialised skills. | Core development outsourced. |
| Individualism. | Different time zones. |
| Uncertainty avoidance. | Shared office hours and office space. |
| Masculinity (assertiveness, competitiveness). | **What** |
| Lack of domain knowledge. | Mobile ICT. |
| Vendor power. | Financial services industry. |
| Business knowledge of vendor. | Application security and privacy issues. |
| Vendor developers' lack of knowledge of the business section. | Telecommunications. |
| Vendor technical skill in IT . | Range of products and services. |
| *Client* | Generic versus custom solutions. |
| Busy schedule. | Payment and expense management solutions. |
| Lack of confidence. | Travel solutions. |
| Lack of motivation. | **Other** |
| Attitude towards relevance of technology. | Strategy. |
| Attitude towards ability to contribute to IT decisions. | Lack of funds. |
| *Relationships* | End users technical infrastructure and working environment. |
| Perception gap between end-users and developers . | Company size . |
| Lack of correlation between analysts' rating of user participation and users' self-rating of involvement. | Open source community. |
| Developing world with abstractive culture. | Modular architecture. |
| Unavailable client. | Modules split between sites (to maximise use of available resources). |
| Lack of common language reference frame. | Customers are individuals through to multi-nationals. |
| Age-gender inequalities. | Roll-out to series of markets / multiple sites in staged approach. |
| Relative positions of parties. | Large-scale, long-term outsourcing. |
| Differences in vocabulary and terminology. | Requirements certainty. |
| Alignment between roles (users, analysts, , developers, testers). | |
| Small close-knit management group. | |

Figure 2: Elements of dimensions

**Strategy** Remove as high level driver for establishing objectives.

**Lack of funds** Unclear at this point. Possibly high level driver of strategy i.e. not a contextual element.

**End user environments** We hypothesise that the underlying factors relate to logistical constraints (*how*) and culture (*who*).

**Company size** We hypothesise that the underlying factors relate to culture (*who*), and spacial and temporal separation (*where*).

**Open source community** This implies a way of viewing things (culture) and separation in space and/or time. We again hypothesise that it is not 'open source community' that affects outcomes, but rather the implied culture and separation.

**Modular architecture** This certainly may be a fixed (contextual) element for some projects. We first check the available dimensions for possible suitability, and realise that, as the factor is connected to the nature of the software product, we should examine the (what) dimension for a possible fit. At present, we identified 'application subject area' (chemical or medical equipment) and 'product relationship' (stand-alone, embedded or middleware) as structural elements. 'Modular architecture' does not fit either. We suggest that a third structural element, 'product structure', might be appropriate. Deeper thought is required.

**Modules split between sites** This is related to 'Module architecture' and represents a possible solution, so we remove it.

**Customers individuals to multi-nationals** As this relates to customer demographic, we hypothesise it is a secondary factor, with underlying factor relating to constraints on both specification and delivery mechanisms (how). It may also be that culture (who) and separation (where) are relevant.

**Roll-out to multiple sites in staged approach** The first part relates to both who and where. The second part is unclear. It may represent a constraint defined by the needs of the customer (how), or it may represent a solution, in which case it is not a contextual element.

**Large-scale outsourcing** This appears to be a solution, but may be viewed as a contextual element if it represents an organisational policy. We hypothesise that, to understand this context, we require a deeper understanding of culture (*who*) and separation (*where*).

**Requirements certainty** This is often cited as affecting outcomes but does not obviously fit into an existing dimension, as currently described. We notice that there may be one of a number of reasons for such uncertainty (also see [12]). One possible reason is that the uncertainty is a result of client-internal processes, for example, if the client understands the application area well, but must wait for a third party before decisions on features can be made. Another possible scenario is that the client is not clear on what is wanted i.e. the product to be built is ill-defined. We make a case for mapping to the what dimension i.e. we must extend this dimension. A third possibility is that the client is weak on decision-making and this infers a mapping to culture i.e. (who). Yet again, we find that we must understand more deeply before we can advise on suit- able practices. For example, a solution of regular client meetings with prototypes will not help if the client is waiting for someone else.

The last entry, 'Requirements certainty', leads us to believe that some context factors mentioned in the literature are, in fact, ambiguous. Without deeper explanation, we simply do not know what is meant.

The use of vague terminology in the software engineering literature has been observed in the area of Global Software Engineering (GSE), where a lack of definition of terms such as 'outsourcing' and 'offshoring' causes confusion in meaning which results in an "inability to judge the applicability and thus transferability of the research into practice" [44]. To address this problem, Šmite et al. have proposed a taxonomy based on the literature and in conjunction with a group of experts. The taxonomy is structured in levels, with key concepts 'location' (on- or off-shore), 'legal entity' (same or different company), 'geographic distance' and 'temporal distance'. Although suitable for many purposes, the taxonomy remains too vague to support identification of suitable practices. For example, there is an assumption that people from the same company have the same culture and this is not necessarily the case. Also, the 'smallest' geographical distance in the terminology is 'close', meaning 'no flights are needed so face-to-face meetings are possible', but this does not address the nuances required for practice selection. For example, are participants in the same room (informal meetings indicated) or in a different building (meetings need to be arranged)?

We note that, for most contextual elements, choosing a single dimension for the element was straightforward. How- ever, when we later examined classifications, we observed some elements classified as 'Where' that we have above claimed should be viewed as secondary. An example is 'Insourcing/outsourcing'. This may be a result of inferring the meaning of the term from the description in the document or it may be an error made during classification. The latter should not be an issue during the full study, as checks will be in place. However, if it is clear from a description that the element really does belong to a single dimension, the lesson is that we must be careful about how we state the element. Our plan was that we would not change terms in any way in order to avoid making assumptions and/or losing nuances of meaning. We may need to reconsider this plan.

## 6. SUMMARY AND FUTURE WORK

We have noted the drive to pursue a more theoretical approach to understanding software practices and have identified the need for an operationalisation of Context as a first step. We have described the pilot for a study that aims to test the suitability of an existing framework for context for application as a theoretical construct. The pilot involved capturing contextual elements from a small selection of the literature into the framework. We wanted to establish if all contextual elements found could be categorised as belonging to a framework dimension (*RQ1*) and to only one framework dimension (*RQ2*). We also hoped to gain some idea of the internal structure of each dimension (*RQ3*).

We found that the majority of the rich set of elements found did, in fact, clearly map to one-and-only-one of the framework dimensions. However, we found some elements that did not. We hypothesise that some of these, for example, 'company size', can be viewed as 'secondary' i.e. it is not the factor as stated that affects practice efficacy, but is rather the elements of the factor, for example, culture and structure. We also hypothesise that each secondary factor can be mapped onto our framework. We also found elements that were ambiguous and so note that deeper investigation is required before these can be mapped. The full study must be adjusted to account for the need to examine secondary and ambiguous factors. The pilot study described was carried out by the first author with ad-hoc checks on progress made by the second author. An element of subjectivity has been introduced. In mitigation, all steps have been documented. The pilot involved studying only three studies from one of the three intended sources. It may be that different kinds of factor will emerge during the full study. This clearly represents a limitation on the findings presented in this paper. However, the pilot resulted in a large number of contextual elements and exposed a set of issues, and these provide some confidence that the framework represents a reasonable first step and was successful as a pilot.

Our contributions are the pursuing of a theoretical approach to understanding software context, the initial establishment and evaluation of a construct for context and the exposure of a lack of clarity of meaning in many 'contexts' currently applied as factors for estimating, for example, project effort.